\documentclass{article}

\usepackage[a4paper, top=3cm, bottom=3cm]{geometry}
\usepackage{natbib}
\usepackage{hyperref}
\usepackage{graphicx}
\newcommand{\Subproblem}[3]{\noindent\textsc{Subproblem.} \normalfont\itshape (#1)
  \newline\noindent \textsc{Input:} #2
  \newline\noindent \textsc{Task:} #3 \normalfont \newline}

\begin{document}

\title{PopIns: population-scale detection of novel sequence insertions}
\author{Birte Kehr\,$^{1,}$\footnote{to whom correspondence should be addressed}\ , P\'all~Melsted\,$^{1,2}$ and Bjarni~V.~Halld\'orsson\,$^{1,3}$}
\date{\small$^{1}$deCODE genetics/Amgen, Reykjav\'ik, Iceland\\
$^{2}$Faculty of Industrial Engineering, Mechanical Engineering and Computer Science, University of Iceland, Reykjav\'ik, Iceland\\
$^{3}$Institute of Biomedical and Neural Engineering, Reykjav\'ik University, Reykjav\'ik, Iceland}

\maketitle

\begin{abstract}

The detection of genomic structural variation (SV) has advanced tremendously in recent years due to progress in high-throughput sequencing technologies.
Novel sequence insertions, insertions without similarity to a human reference genome, have received less attention than other types of SVs due to the computational challenges in their detection from short read sequencing data, which inherently involves de novo assembly.
De novo assembly is not only computationally challenging, but also requires high-quality data.
While the reads from a single individual may not always meet this requirement, using reads from multiple individuals can increase power to detect novel insertions.

We have developed the program \emph{PopIns}, which can discover and characterize non-reference insertions of 100\,bp or longer on a population scale.
In this paper, we describe the approach we implemented in \emph{PopIns}.
It takes as input a reads-to-reference alignment, assembles unaligned reads using a standard assembly tool, merges the contigs of different individuals into high-confidence sequences, anchors the merged sequences into the reference genome, and finally genotypes all individuals for the discovered insertions.
Our tests on simulated data indicate that the merging step greatly improves the quality and reliability of predicted insertions and that \emph{PopIns} shows significantly better recall and precision than the recent tool \emph{MindTheGap}.
Preliminary results on a data set of 305 Icelanders demonstrate the practicality of the new approach.

The source code of \emph{PopIns} is available from \href{http://github.com/bkehr/popins}{http://github.com/bkehr/popins}.

\end{abstract}

\section{Introduction}

The latest version of the human reference genome~\citep{Venter2001,Lander2001}, GRCh38, is of a remarkable quality.
However, the sequence of a single individual is inherently different from the reference due to sequence diversity.
Some sequences are missing in the reference as they are not present in the individuals from whom the reference was constructed.
Alternate haplotypes have been added to the reference genome~\citep{Horton2008} to account for highly variable regions, but they cover only a small part of the variation.
The variable regions are of great biological and medical interest since their sequence diversity is known to affect phenotypes including numerous diseases~\citep{Stankiewicz2010,Conrad2010}.
Thus, the characterization of differences to the reference genome is a major task.

Differences between human genomes include single nucleotide polymorphisms (SNPs), small indels, and structural variants (SVs).
One type of SVs, which affect a larger piece of sequence than indels, are insertions.
Insertions can be further classified into duplications and novel sequence insertions.
Duplications are insertions of sequence also present elsewhere in the genome, e.\,g.~ mobile elements.
The focus of this work is on novel sequence insertions, insertions of unique sequence that is not similar to other regions of the reference genome.
The evolutionary origin of a novel sequence insertion may be explained by two types of events.
On the one hand, it may be an addition of sequence to the genome of a sequenced individual, e.\,g.~lateral transfer or viral genome insertion.
On the other hand, it may as well be a deletion of sequence in the individuals used to contruct the reference.

In order to successfully associate genetic differences with phenotypes, large numbers of individuals are necessary.
With the enormous improvements in sequencing technology, recent years have seen a marked increase in large-scale efforts that aim at discovering variation at a population level, e.g. the HapMap project~\citep{HapMap}, the 1000 genomes projects~\citep{1000genomes,1000genomes2012}, and the Genome of the Netherlands project~\citep{Boomsma2014}.
The achievements of these efforts have been the characterization of a great number of SNPs, indels, and deletions, but comparatively fewer novel insertions.
For example, Mills et al.~\citep{1000genomesCNV} reported only 128 novel insertions in contrast to 22025 deletions in their release set.

One reason for a smaller number of insertions discovered is the fact that their detection from short read sequencing data is challenging~\citep{1000genomesCNV}.
Unlike detection of other types of variation, insertion detection requires a de novo assembly.
Typical genotype callers, such as \emph{GATK}~\citep{gatk,gatkng} or \emph{FreeBayes}~\citep{freebayes}, use only reference-aligned read pairs and, therefore, are not suitable for calling insertions longer than the reads.
Sequencing technologies that yield longer reads~\citep{pacbioClosing} show promise in simplifying insertion detection~\citep{pacbioComplexity}, but they are still not commonplace nor cost-effective on a large scale.
Thus, insertions remain one of the most challenging types of variation to detect.

Strategies for detecting insertions incorporate either a local assembly of unknown sequence not present in the reference genome~\citep{MindTheGap,BasilAnise,novelSeq,tigra} or a whole-genome assembly~\citep{velvet,masurca,spades,soapDenovo,allpaths}.
A whole-genome assembly needs to be followed by a comparison step of the assembled contigs to the reference genome for identifying the insertions.
In the local assembly strategy, the positions of the insertions need to be identified either before or after assembly.
For example, the strategy implemented in the program \emph{MindTheGap}~\citep{MindTheGap} first identifies candidate insertion sites without a read alignment before initiating local assemblies, whereas the \emph{NovelSeq} approach~\citep{novelSeq} first assembles unaligned reads before anchoring them in the genome.

Both the whole-genome and the local assembly strategies face difficulties when integrating the results of many individuals.
This issue is most pronounced in \emph{de novo} whole-genome assembly strategies and, thus, has previously been addressed.
Parrish et al.~\citep{parrish2013genome} suggested a reference-guided whole-genome assembly approach that makes the results of several samples more compatible.
Iqbal et al.~\citep{cortex} developed \emph{Cortex}, a program that rigorously assembles the whole genomes of several individuals at the same time based on colored de-Bruijn graphs.
However, the tests in the \emph{Cortex} paper were limited to relatively few individuals or to pooled data.
Furthermore, all whole-genome assembly strategies commonly suffer from a considerable demand for computational resources.

In addition, the assembly problem demands high-coverage data~\citep{Miller2010,Zerbino2012}.
Assemblies from low-coverage data are typically incomplete, i.e. fragmented and with significant portions of the sequences missing~\citep{Alkan2011}.
Hence, the application of any of the mentioned approaches on a data set from a single individual sequenced at insufficient coverage, results in a largely incomplete set of insertions.
Polymorphic insertions with low-frequency in a population are particularly hard to detect as they are likely to appear only at heterozygous loci.
Additionally, incomplete assemblies lead to greater difficulties in comparing and integrating sets of insertions across multiple individuals.
If not carefully considered at the population level, all this can add to an underestimation of allele frequencies, less power to detect rare insertions, and may eventually impede association with phenotypes.

Despite the caveats mentioned, the task of analyzing large numbers of individuals can also aid in the detection of insertions.
Insertions that occur within many individuals have an increased total coverage across the whole data set.
If used in the assembly step, this may reduce fragmentation and fill in gaps of the insertion sequences.
The larger the number of individuals, the more likely it is that we can capture low-frequency insertions.
Thus, instead of merging insertions detected from many individuals, we can take advantage of all individuals during the detection of insertions.

We have developed an approach for characterizing insertions across a large number of individuals simultaneously using a local assembly strategy.
We start by assembling per individual reads that do not align to the reference genome.
Subsequently, we merge the assemblies into a multi-individual contig set of higher quality.
Each contig in this set is then placed into the reference genome using read-pair and split-read information.
Finally, we propose a genotyping procedure that determines for each individual the number of copies it carries of an insertion in its diploid genome.

We have implemented the approach in a program called \emph{PopIns}.
Our tests on simulated data indicate that merging of single-individual assemblies increases the quality of insertion sequences by more than 20\,\%.
A comparison to \emph{MindTheGap}~\citep{MindTheGap} confirms that we greatly benefit in recall from the merging step and that our approach is precise.
An additional test on data from 305 whole genomes obtained with Illumina sequencers demonstrates its practicality on real data.

\section{Problem formulation and approach}

\begin{figure*}[!tpb]
\centerline{\includegraphics[width=0.75\linewidth]{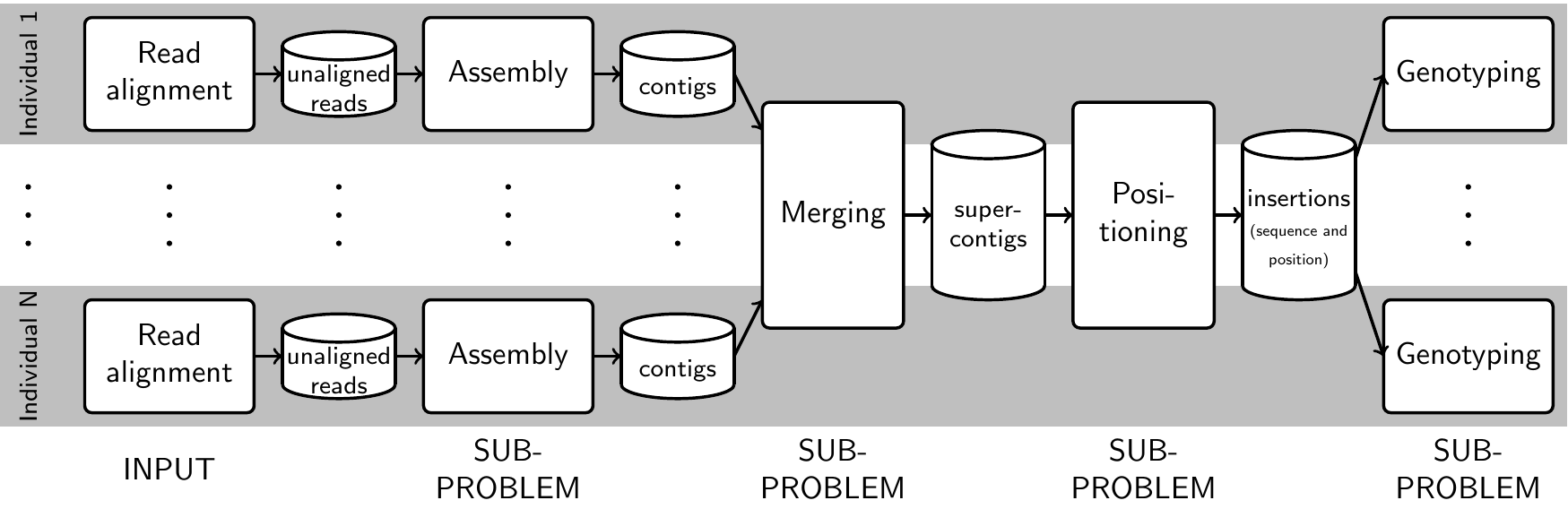}}
\caption{Overview of our approach. Starting from a read alignment, we break up the insertion detection problem for many individuals into four subproblems: assembly, merging, positioning, and genotyping.}
\label{fig:subproblems}
\end{figure*}

Given reference-aligned paired-end sequencing data from many individuals of the same species, the problem we deal with in this paper is to identify a set of long novel sequence insertions with respect to the reference genome and to determine the genotypes of all individuals for each insertion in the set.

\subsection{Polymorphic long novel sequence insertions}

An insertion in a multi-individual data set is fully defined by three attributes: a position, a sequence, and a genotype of each individual.
The sequence of a \emph{novel sequence insertion} is not similar to any part of the reference genome.
By a \emph{long} insertion we here refer to a sequence longer than the sequencing reads of the given data set.
An insertion is \emph{polymorphic}, if it is present in one or two copies of an individual's diploid genome and in one or more individuals within the given data set at a frequency below 100\,\%.
The \emph{genotype} of a single diploid individual determines the number of copies the individual carries of a polymorphic insertion.
Possible genotypes are \emph{homozygous refererence} (zero copies), \emph{heterozygous} (one copy), or \emph{homozygous insertion} (two copies).

\subsection{A local assembly approach for a single individual}

Before we extend to multiple individuals, we consider the problem where a sequencing data set is given for a single individual.
We assume that this data set has been aligned to a reference genome and refer to high-quality reads without alignment or with very low alignment scores as the set of unaligned reads.
We define three separate subproblems, each addressing one of the three attributes of an insertion: the sequence, the position, and the genotype.
Consecutive solutions to these three subproblems form a local assembly approach for a single individual.

\medskip
\Subproblem{Assembly}
  {The set of unaligned reads from a sequencing data set of a single individual.}
  {Reconstruct a set of contigs, representing candidate sequences of insertions.}

\noindent We note that this problem is a classical genome assembly problem~\citep{Miller2010}.

\medskip
\Subproblem{Positioning}
  {An alignment of reads to the reference genome, a set of contigs, and an alignment of the unaligned reads to these contigs.}
  {Determine the position on the reference genome where the contig is inserted.}

\noindent We assume a solution to this problem to provide exactly one position for each contig although in practice more than one position is possible.

\medskip
\Subproblem{Genotyping}
  {A contig, a position, and an individual's reads that align to the contig and to the reference near the position.}
  {Classify the individual as homozygote reference, heterozygote, or homozygote insertion.}

\subsection{A local assembly approach for multiple individuals}

We now extend the single-individual problem to the case where aligned read data of multiple individuals is given.
We could solve the above three subproblems for each individual separately.
However, in order to obtain a single set of insertions for all individuals, an additional merging step is necessary.

If we merge the sets of insertions after solving all three subproblems, we do not benefit from multi-individual data and expect many false negative insertions due to low coverage.
When genotyping is done after merging the sets, we can discover an insertion in all individuals even though the detection failed in some individuals.
However, this leads to extra computational cost when solving the positioning subproblem for similar contigs of multiple individuals.
In addition, it is challenging to identify insertions in different individuals as the same insertion if the positions are inaccurate.

Therefore, we add a merging step after solving the assembly subproblem but before solving the positioning subproblem.
This step merges sets of contigs into a single contig, which we call a \emph{supercontig}.
We say that a supercontig represents a contig if the full length of the contig aligns to the supercontig.

\medskip
\Subproblem{Merging}
  {Given multiple sets of contigs, where each set corresponds to one individual.}
  {Reconstruct a set of supercontigs that fulfils the following two conditions:}
  \textit{
  \vspace{-5mm}
  \begin{enumerate}
  \item Every input contig is represented by a supercontig in the set.
  \item Any two supercontigs in the set cannot be further merged.
  \end{enumerate}}

\noindent We note the similarity of this subproblem to a classical assembly problem and discuss the differences in Sect.~\ref{sec:discussion}.

Our approach for detecting insertions in many individuals solves the subproblems assembly, merging, positioning, and genotyping in this order as is illustrated in Fig.~\ref{fig:subproblems}.

\section{Methods}
This section presents our solutions to the four subproblems and describes the simulation of a test data set.
Details of our solutions to the subproblems can be found in corresponding subsections of Sect.~S1 in the supplementary material.

\subsection{Assembly subproblem}

As noted above, the assembly subproblem is a classical genome assembly problem.
Careful work has been put into the development of tools for genome assembly from short-read data, among them \emph{Velvet}~\citep{velvet}, \emph{SPAdes}~\citep{spades}, and \emph{MaSuRCA}~\citep{masurca}.
All these tools compute a set of contigs when given an individual's set of unaligned reads and, thus, can be applied to solve the assembly subproblem within our approach.
We integrated \emph{Velvet} into our implementation (see supplementary Sect.~S1.1).

\subsection{Merging subproblem}

\begin{figure*}[!tpb]
\centerline{\includegraphics[width=0.6\linewidth]{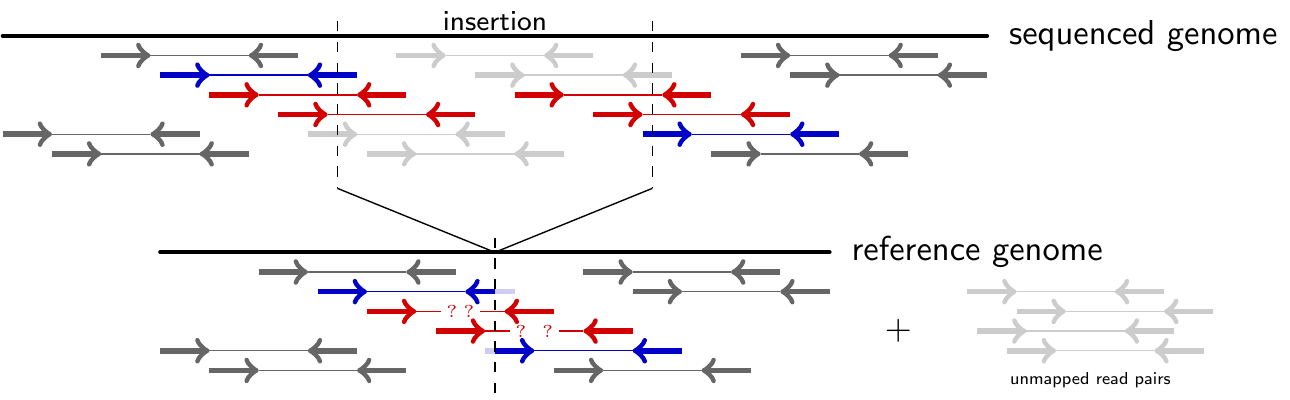}}
\caption{The alignment of reads near a novel sequence insertion to the genome of a homozygous carrier of the insertion (top) and to the reference genome (bottom).
Black read pairs correctly align to the reference genome, one end of red read pairs does not align to the reference, one end of each blue read pair is a split-read, and grey read pairs do not align to the reference.}
\label{fig:insertion}
\end{figure*}

For the merging subproblem, we develop a practical solution that scales to hundreds of contigs (totaling to a length of hundreds of kilobases) from each of thousands of individuals.
Formally, the inputs to the merging subproblem are sets of contigs $C_1, \dots, C_n$, one for each of $n > 1$ individuals.
Let $\mathcal{C} = \bigcup_i C_i$.
Our objective is to merge the contigs in $\mathcal{C}$ into a supercontig set $S$ of minimal size such that for each contig $c \in \mathcal{C}$ a supercontig $s \in S$ exists that represents $c$.
We say that a supercontig $s$ \emph{represents} a contig $c$ if $c$ aligns to a substring of $s$ with an error rate of at most $\varepsilon$ (default $\varepsilon = 0.05$), where the error rate is calculated as the edit distance divided by the length of $c$.

We apply a two-step approach to solve the merging subproblem.
It is based on the assumption that, in a minimal set of supercontigs, any two contigs $c_p$ and $c_q$ represented by the same supercontig $s \in S$ are connected by a path of contigs $c_p = c_{i_1}, \dots, c_{i_r} = c_q$ such that any two adjacent contigs $c_{i_x}$ and $c_{i_{x+1}}$, $i_1 \leq i_x < i_r$ locally align to each other.
Following this assumption, our first step aligns and partitions the original set of contigs $c \in \mathcal{C}$ into new sets $D_1, \dots, D_k$, such that any two contigs that are elements of the same set $D_j$, $1 \leq j \leq k$ are connected by a path and any two contigs that are connected by a path are elements of the same set.
The second step determines the supercontigs for each set $D_1, \dots, D_k$.
In the ideal case, all contigs in a set $D_j$, $1 \leq j \leq k$, originate from the same insertion and can be merged into exactly one supercontig.

In order to partition the contigs in $\mathcal{C}$ into sets $D_1 \dots D_k$, we apply a union-find data structure on sets of contigs.
We iteratively add the contigs in $\mathcal{C}$ to an initially empty union-find instance $\mathcal{D}$ such that, in the end, $\mathcal{D}$ represents the sets $D_1 \dots D_k$.
We unify two sets whenever we find an alignment between the current contig and any contig in each of the two sets.
Instead of aligning all pairs of contigs with dynamic programming, we pre-filter for potential alignments with the fully sensitive k-mer counting algorithm \emph{SWIFT}~\citep{swift}.
Supplementary Sect.~S1.2.1 gives further details on our partitioning step.

Subsequently we construct the supercontigs, separately for each set $D_1, \dots, D_k \in \mathcal{D}$.
This construction requires a multiple sequence alignment, which is known to be an NP-hard problem~\citep{Wang1994}.
Thus, we apply an iterative approach similar to progressive alignment~\citep{Feng1987}.
In order to explore the possibility that the contigs do not assemble into a single contiguous supercontig, we use a graph data structure.
Graph nodes are labeled with substrings of the contigs; the graph edges are directed and represent adjacencies of the contig substrings.
After adding all contigs to the graph, we obtain supercontigs by following all paths through the graph.
Supplementary Sect.~S1.2.2 provides the details on how we iteratively add contigs to the graph.

\subsection{Positioning subproblem}

Our solution to the positioning subproblem uses information from read pairs as well as split reads.
For a given insertion, we define an \emph{anchoring read pair} as a pair of read ends where one end aligns to the reference genome and the other end to the insertion.
A \emph{split read} is a read end without a full-length alignment to the reference genome or insertion, but with a position, the \emph{split position}, that divides the read end into a prefix aligning to the reference genome and a suffix aligning to the insertion, or a prefix aligning to the insertion and a suffix aligning to the reference genome.
Figure~\ref{fig:insertion} shows how anchoring read pairs and split reads of an insertion align to the reference genome and to the genome of a homozygous carrier of the insertion.
For anchoring read pairs, the orientations of the read ends in the alignments distinguish four different scenarios at a single genomic location (see Fig.~S1 in the supplementary material).
For distinguishing the four scenarios with split reads, it additionaly matters whether the prefix or suffix aligns to the reference.

Our positioning approach considers the two ends of each contig separately.
This allows the detection of more complex events than simple insertions, for example an insertion of novel sequence that is followed by a mobile element.
For each contig end, we first use anchoring read pairs to find the most probable locations in the reference genome and then split reads to determine the exact insertion positions at these locations.

We determine potential locations of a contig end by clustering anchoring read pairs using a greedy approach that scans linearly over the anchoring read pairs sorted by position to identify sets of read pairs that pairwisely support the same insertion.
Afterwards, we compute an anchoring score between 0 and 1 for each location that takes into account all alternative locations. 
See supplementary Sect.~S1.3.1 for a more detailed description of the cluster and score computation.
We keep only locations that have an anchoring score above a fixed threshold $\alpha$ (default $\alpha = 0.3$) and those that are shorter than twice the maximum allowed insert size (default $d_{max}=800$) as all others are unlikely to be correct.

For each of the remaining locations, we search for the exact insertion position using split reads.
While the anchoring read pairs can give a first estimate of this position, split reads provide exact predictions of the position at base pair resolution.
We collect potential split reads, align them using the split alignment function from the SeqAn C++ library~\citep{seqan}, and determine the insertion position from the whole set of split-aligned reads.
See supplementary Sect.~S1.3.2 for details.

\begin{table*}[t]
\begin{center}
\caption{Results of all steps of \emph{PopIns} on simulated data.
}
\label{tab:sim-results}
\begin{tabular}{l@{\qquad}r@{\quad}r@{\qquad}r@{\quad}r@{\qquad}r@{\quad}}
\hline\rule{-3pt}{12pt}
  & \multicolumn{2}{c@{\qquad}}{full sequence/} & \multicolumn{2}{c@{\quad}}{partial sequence/} & \multicolumn{1}{c}{total} \\
  & \multicolumn{2}{c@{\qquad}}{both ends}      & \multicolumn{2}{c@{\quad}}{at least 1 end}    & \multicolumn{1}{c}{simulated} \\[2pt]
\hline\rule{-3pt}{12pt}
Assembled (single individual)            & 62.3\,\% & 4243 & 75.5\,\% & 5111 & 6768 \\
Supercontigs after merging               &   85\,\% &   85 &   90\,\% &   90 &  100 \\
Locations (from anchoring read pairs)    &   90\,\% &   90 &   90\,\% &   90 &  100 \\
Positioned (with split reads)            &   84\,\% &   84 &   90\,\% &   90 &  100 \\
Correctly genotyped                      & 85.4\,\% & 5783 & 90.5\,\% & 6122 & 6768 \\[2pt]
\hline
\end{tabular}
\end{center}
\footnotesize
The column \emph{full sequence/both ends} considers only those predictions where no more than 4\,bp are missing from one of the insertion sequence's ends or both ends are positioned/genotyped.
The column \emph{partial sequence/at least 1 end} includes also predictions where one end of the insertion sequence is truncated or only one end is positioned or correctly genotyped.
The numbers include one insertion sequence that is split into two supercontigs.
\end{table*}

\subsection{Genotyping subproblem}
      
Given the sequence of an insertion, its position on the reference genome, and an individual's sequence read data.
Our solution to the genotyping subproblem constructs the sequence of the two possible alleles at the insertion position $R$ (reference) and $A$ (alternate, i.e. insertion) in a window of size $2w$ (default $w = 50$), re-aligns the set of reads $\mathcal{R}$ to both alleles using dynamic programming (see supplementary Sect.~S1.4 for details), and computes a relative likelihood for an individual to carry the insertion on 0, 1 or 2 of his chromosomal copies based on alignment scores.
As above, we consider both ends of the insertions separately.

We compute the likelihood based on the alignment scores of all reads $r \in \mathcal{R}$.
Given a read $r$ aligned to one of the two allele sequences $S \in \{A,R\}$ with score $s_S$, we assume $P(r|S) \sim e^{s_S}$.
As we are only interested in relative likelihoods of $R$ and $A$, we compute 
$$P(r|S) \sim \frac{e^{s_S}}{e^{s_A}+e^{s_R}}$$
where $s_A$ and $s_R$ are the alignment scores to $A$ and $R$, respectively.
To ensure that a single read does not have too large of an impact in the joint computation of the genotype likelihood described below, we bound this relative likelihood from below by a small constant $c$ (default $c = 0.0001$).

If we assume that reads are sampled from the two alleles of an individual with equal probability, we get the genotype likelihoods:
$$P(r|R,R) \sim     \frac{e^{s_R}}{e^{s_A}+e^{s_R}}$$
$$P(r|R,A) \sim     \frac{1}{2}\frac{e^{s_R}}{e^{s_A}+e^{s_R}} + \frac{1}{2}\frac{e^{s_A}}{e^{s_A}+e^{s_R}}=\frac{1}{2}$$
$$P(r|A,A) \sim     \frac{e^{s_A}}{e^{s_A}+e^{s_R}}$$
Under the assumption that the reads in $\mathcal{R}$ are independent, we get the likelihoods for an individual's genotype $(S_1,S_2) \in \{A,R\}\times\{A,R\}$:
$$P(R|S_1,S_2) \sim \Pi_{r \in \mathcal{R}} P(r|S_1,S_2)\ .$$
Finally, we pick the genotype that has the highest likelihood as our prediction and require this likelihood to be above 0.5 by default.

\subsection{Test data sets}

For performance evaluation, we simulated a data set with polymorphic insertions at random positions.
Furthermore, we selected a real data set of 305 Icelanders as described below and data from the individual NA12878 as described in Sect.~S2.3 of the supplementary material.
Supplementary Sect.~S2 describes how we process the sequencing reads from all data sets to obtain a set of unaligned reads as input to the subproblems.

\subsubsection{Simulation of data.}

Our simulation is based on human chromosome 18, GRCh37 (hg19).
We simulated 100 insertions that occur at different frequencies in 100 diploid individuals.
In order to get realistic insertion sequences, we deleted 100 randomly selected regions from the reference chromosome using the resulting sequence as our reference and keeping the deleted sequences as insertions.
We chose the lengths of these insertions according to an exponential distribution with a mean of 100 and always added 100 bps.
Next, we uniformly sampled a frequency between 0 and 1 for each insertion and added the insertion sequence back into 200 copies of the modified reference at the assigned frequency.
The 200 copies of the modified reference each equipped with a subset of the insertions represent a set of 200 simulated haplotypes.

For each haplotype, we simulated sequencing reads at a coverage of $13\times$.
From the Mason read simulation package, version 2.0~\citep{mason}, we used the variator tool to add SNPs and small indels and the simulator tool to generate 5 Million 101 bp paired-end Illumina reads per haplotype.
Finally, we merged the sequencing reads of haplotype pairs to obtain data at a coverage of $26\times$ for 100 diploid individuals.
Our data set created with this simulation procedure has an average of 32.19 heterozygous and 35.49 homozygous insertions per individual.

\subsubsection{Real data of 305 individuals.}

To test our approach on real sequencing reads, we selected the data of 305 Icelanders sequenced at deCODE genetics \citep{decodeSequencing}.
These data were generated by Illumina HiSeq instruments in paired-end mode with a read length of 101\,bp and an average insert size of 402\,bp.
The number of reads yielded sequencing coverages between $8\times$ and $45\times$ (average $24\times$).

\section{Results}

We implemented our solutions to the merging, positioning, and genotyping subproblems in a program called \emph{PopIns} using the SeqAn C++ library~\citep{seqan}.
We assess the performance of \emph{PopIns} on the simulated data set, compare its results to that of the program \emph{MindTheGap}~\citep{MindTheGap}, and demonstrate its practicality on the data of 305 Icelanders.
The supplementary material describes in Sect.~S3.3.1 and Table~S2 an additional experiment on a single Icelandic individual that quantifies the benefit from the merging step.
Further, Sect.~S3.3.2 to~S3.3.4 compare the performance of \emph{PopIns} on data from the NA12878 individual to insertions validated by \citep{kim2013} and insertions predicted in the 1000 genomes project~\citep{1000genomes}.

\subsection{Performance on simulated data}

Table~\ref{tab:sim-results} summarizes the results of \emph{PopIns} on our simulated data set.
The unaligned reads assemble into an average of 51.45 contigs per individual, which align to 75.5\,\% of the 6768 simulated insertion sequences using \emph{Stellar}~\citep{stellar} with a minimal length of 50 and a maximal error rate of 5\,\%.
After merging, the supercontigs align to 90\,\% of the 100 simulated insertion sequences.
With anchoring read pairs we find approximate locations for all supercontig ends.
The anchoring score always points to the correct location if multiple locations are suggested.
For 85 supercontigs, \emph{PopIns} finds an exact insertion position for both ends using split reads.
For the remaining five supercontigs, \emph{PopIns} finds an exact insertion position only for one end.
These supercontigs turn out to be truncated with respect to the insertion sequence by 5 or more base pairs at the other end.
Finally, \emph{PopIns} genotypes 172 ends of insertions in all individuals correctly, reports a false position as heterozygous in 23 individuals, and discards 6 false positions in all individuals by typing them as homozygous reference.
In relation to the 6768 single-individual insertions, this yields a recall of 85.4\,\% and a precision of 99\,\%  after genotyping when counting only those insertions where both ends are fully characterized.
Supplementary Sect.~S3.1 provides further details of our evaluation on simulated data.

Furthermore, we made an experiment of pooling the unaligned reads of all individuals before assembly with velvet (using the same parameters except for an extended coverage window of $[2,10000]$).
This yields only 14 contigs, a small number possibly reflecting assumptions made by velvet (e.g. on even coverage), a program that was developed for whole-genome assembly of a single individual.

\subsection{Comparison to \emph{MindTheGap}}

We use our simulated data set for comparison of \emph{PopIns} to \emph{MindTheGap}~\citep{MindTheGap}.
We ran the index, find, and fill commands of \emph{MindTheGap} with default parameters for each individual separately and selected all predicted insertions longer than 99 bps.
In contrast to \emph{PopIns}, \emph{MindTheGap} first identifies potential insertion positions and then assembles their sequences.
It reports each insertion as heterozygous or homozygous, which we interpret as the individuals genotype in our comparison.
In many cases, MindTheGap reports several sequences per position.
Following Rizk et al.~\citep{MindTheGap}, we count only one sequence per position for calculating the precision, which gives a favorable representation of results of \emph{MindTheGap}.
We aligned all sequences to the set of simulated insertions (using \emph{Stellar} as above) and counted a simulated insertion as found if 90\,\% or more of the sequence aligned to a predicted sequence.

Table~\ref{tab:sim-results-comparison} shows the recall and precision of \emph{PopIns} in comparison to \emph{MindTheGap}.
\emph{PopIns} clearly outperforms \emph{MindTheGap} in terms of recall.
Further, the predictions of \emph{PopIns} are more reliable in terms of precision, even though we count only one sequence per position for \emph{MindTheGap}.
We observed a much smaller difference in recall and precision between homozygous and heterozygous insertions for \emph{PopIns} than for \emph{MindTheGap} (data not shown).

We count predictions per individual and consider an insertion as true positive if the position, sequence, and genotype is correct.

\subsection{Running times}

\emph{PopIns} and \emph{MindTheGap} were run exclusively on Cisco 4\,GHz machines with 16\,GB of memory in a compute cluster.
The total CPU time used by \emph{PopIns} to finish all steps from assembly to genotyping for all individuals was 3:30\,h.
In comparison, \emph{MindTheGap} used 35:54\,h in total for the index, find, and fill steps of all individuals.
Thus, the computations of \emph{PopIns} take $\sim$10 times less time.
The difference is smaller between the actual walltimes that include I/O operations but difficult to quantify as the walltimes depended heavily on the total cluster load.
Supplementary Sect.~S3.2 lists factors that influence running time in \emph{PopIns}.

\begin{table}[t]
\caption{Comparison of \emph{PopIns} and \emph{MindTheGap} on simulated data.}
\label{tab:sim-results-comparison}
\begin{center}\begin{tabular}{l@{\qquad}r@{\quad}r}
\hline\rule{-3pt}{12pt}
                   & Precision & Recall \\[2pt]
\hline\rule{-3pt}{12pt}
\emph{PopIns}      & 99.0\,\%  & 85.4\,\% \\
\emph{MindTheGap}  & 81.9\,\%  & 69.5\,\% \\[2pt]
\hline
\end{tabular}\end{center}
\end{table}

\subsection{Performance on data of 305 human individuals}

We tested the practicality of \emph{PopIns} on the data of 305 Icelanders.
After processing the data as described in Sect.~S2.2 of the supplementary material, we obtain an average of 35531 unaligned reads per individual.
The reads assemble into an average of 691 contigs per individual with an average N50 of 334 and totaling to an average length of 209\,kbp.
The merging step reduces the set of 210892 contigs to 8437 supercontigs including 6141 contigs that are unique to one individual.
We identify two sets of individuals (of size four and six) that have extra contigs, which are likely to originate from bacterial species (contamination).
We exclude supercontigs that were only found in the contaminated individuals and thereby reduce the set of supercontigs to a size of 2226 of which only 401 are unique to one individual.

After aligning the unaligned reads to the set of supercontigs, anchoring read pairs suggest 263732 locations for the supercontig ends, of which 2686 have an anchoring score above 0.3 and are supported by more than one anchoring read pair.
For making a manual evaulation of the predicted insertions possible, we restricted further analysis to chromosome 18.
Of all locations, 3968 fall on chromosome 18, of which 66 have an anchoring score above 0.3 and are supported by more than one anchoring read pair.
Among the 66 locations, 46 can be paired into 23 records that explain the two ends of one supercontig.
At 37 of the 66 locations we can determine the exact position from an unambiguous set of split reads, including 14 pairs for the two ends of one supercontig.
At 36 of the 37 positions, the genotyping algorithm determines at least one individual to be a carrier of the insertion.
The insertion frequencies in the 305 individuals range from 0.5\,\% to 100\,\%.
Fig.~S2 in the supplementary material displays example read alignments for three individuals around one of the insertions, which visualizes that the data strongly supports the different genotype calls.

\section{Discussion}\label{sec:discussion}

We have introduced \emph{PopIns}, a method for discovering and genotyping novel sequence insertions.
\emph{PopIns} takes advantage of the information provided by many individuals, increasing the quality of the insertion sequences and significantly improving on our ability to determine the correct insertion position.
Our local assembly approach reduces computational requirements as compared to whole-genome assembly~\citep{cortex}.
On the downside, our approach depends on a read alignment and, thus, will be biased against the reference.

The major novelty of our approach is the addition of a merging step to a local assembly strategy.
The merging subproblem is \emph{per se} an assembly problem but significantly different from the classical genome assembly problem.
The input sequences of the merging subproblem are themselves assembled contigs and, hence, can contain artifacts from misassemblies.
In addition, they vary in length and we expect them in the best case not to be much shorter than the supercontigs.
Similar to transcriptome assembly, the coverage is uneven depending on the number of individuals that are carriers of an insertion and, finally, more than two haplotypes are possible for each insertion locus as the contigs originate from many individuals.

Nevertheless, our solution to the merging subproblem is similar to the overlap-layout-consensus (OLC) approach for genome assembly~\citep{Miller2010}.
The use of the union-find data structure is similar to the graph used in the overlap phase of OLC approaches; our sets of contigs correspond to connected components in this graph.
Our approach differs from the OLC approach in the layout phase by allowing for branching components when constructing supercontigs.

When finding insertion positions, \emph{PopIns} greedily clusters anchoring read pairs by location, while other SV detection methods solve a maximum clique problem~\citep{delly,clever}.
Our approach can lead to very long intervals for a single location.
But since we cluster the anchoring read pairs only per contig and not over the whole data set, we observe only very few abnormally long intervals.
We discard these applying a length threshold as they are unlikely to lead to a clear insertion position.

In our evaluation, we were not always able to find the insertion position for both contig ends, which can have several reasons.
If we find a single location with anchoring read pairs but no clear position with split reads, the contig is likely not to contain the whole insertion.
The split alignment algorithm we used penalizes all gaps in the reference.
Allowing for a large gap in both the reference and the reads may help in narrowing down the position of these insertions.
Another reason may be non-unique sequence being inserted together with the novel sequence.
The set of unaligned reads will not assemble into contigs of non-unique sequence (e.g. mobile elements), thus, these are missing in our approach.
In many cases, this leads to many low-scoring locations suggested by read pairs that anchor to known occurrences of the repeated sequence.
Finally, we observe read pairs connecting several contigs, suggesting insertions of novel sequence interspersed with non-unique sequence.
An additional scaffolding step of supercontigs would be necessary to fully characterize these cases.

In many cases we observe the same short sequence repeated at the two ends of an insertion (often referred to as target site duplications).
If the repeated sequences become too long, our genotyping approach has difficulties in distinguishing the reference allele from the alternate allele.
This may potentially be improved by focusing the computation on the unique part of the sequence.

Our results on simulated data do not reflect all of these limitations of \emph{PopIns} that are due to the complex structure of real genomic sequences.
Still, we could show the practicality of the approach on real data, where it yields many novel sequence insertions.
Therefore, we are expecting a rich set of polymorphic insertion when applying \emph{PopIns} to a larger number of individuals, which will open up the door to include novel sequence insertions in genome-wide association studies.

\bibliographystyle{natbib}

\bibliography{document}

\end{document}